\documentclass[10pt, journal]{IEEEtran}
\IEEEoverridecommandlockouts
\usepackage{cite}
\usepackage{amsmath,amssymb,amsfonts,mathtools, bm}
\usepackage{amsthm}
\usepackage{algorithm}
\usepackage{algorithmic}
\usepackage{graphicx}
\usepackage{textcomp}
\usepackage{xcolor,float}
\usepackage{tabularx}
\usepackage{textcomp}
\usepackage{diagbox}
\usepackage{hyperref}
\usepackage{booktabs}

\newtheorem{problem}{Problem}

\usepackage{color,float}
\usepackage{dsfont}
\usepackage{diagbox}
\ifCLASSINFOpdf
\else
\fi

\begin{document}
\title{An Enhanced Dual-Currency VCG Auction Mechanism for Resource Allocation in IoV: A Value of Information Perspective}
\author{Wei Wang,~\IEEEmembership{Student~Member,~IEEE,}
        Nan Cheng,~\IEEEmembership{Senior Member,~IEEE,}
        Conghao Zhou,~\IEEEmembership{Member,~IEEE,}
        Haixia Peng,~\IEEEmembership{Member,~IEEE,}
        Haibo Zhou,~\IEEEmembership{Senior Member,~IEEE,}
        Zhou Su,~\IEEEmembership{Senior Member,~IEEE,}
        Xuemin (Sherman)  Shen,~\IEEEmembership{Fellow, IEEE}
        \IEEEcompsocitemizethanks{\IEEEcompsocthanksitem 
        \IEEEcompsocthanksitem Wei Wang is with the School of Telecommunications Engineering, Xidian University, Xi'an, China (e-mail: weiwang\_2042@stu.xidian.edu.cn).
        \IEEEcompsocthanksitem Nan Cheng (corresponding author) is with the State Key Laboratory of ISN and School of Telecommunications Engineering, Xidian University, Xi'an, China (e-mail: dr.nan.cheng@ieee.org). 
        \IEEEcompsocthanksitem Haixia Peng is with the School of Information and Communications Engineering, Xi'an Jiaotong University, Xi'an, China (e-mail:  haixia.peng@xjtu.edu.cn).
        \IEEEcompsocthanksitem Haibo Zhou is with the School of Electronic Science and Engineering, Nanjing University, Nanjing 210023, China (e-mail: haibozhou@nju.edu.cn).
        \IEEEcompsocthanksitem Zhou Su is with the School of Cyber Science and Engineering, Xi'an Jiaotong University, Xi'an, China  (e-mail: zhousu@ieee.org).
        \IEEEcompsocthanksitem  Conghao Zhou and Xuemin (Sherman) Shen are with the Department of Electrical and Computer Engineering, University of Waterloo, Waterloo, N2L 3G1, Canada (e-mail: c89zhou@uwaterloo.ca; sshen@uwaterloo.ca).}
        }
\maketitle
\begin{abstract}

The Internet of Vehicles (IoV) is undergoing a transformative evolution, enabled by advancements in future 6G network technologies, to support intelligent, highly reliable, and low-latency vehicular services. 
However, the enhanced capabilities of loV have heightened the demands for efficient network resource allocation while simultaneously giving rise to diverse vehicular service requirements.
For network service providers (NSPs), meeting the customized resource-slicing requirements of vehicle service providers (VSPs) while maximizing social welfare has become a significant challenge. This paper proposes an innovative solution by integrating a mean-field multi-agent reinforcement learning (MFMARL) framework with an enhanced Vickrey-Clarke-Groves (VCG) auction mechanism to address the problem of social welfare maximization under the condition of unknown VSP utility functions. The core of this solution is introducing the ``value of information" as a novel monetary metric to estimate the expected benefits of VSPs, thereby ensuring the effective execution of the VCG auction mechanism. MFMARL is employed to optimize resource allocation for social welfare maximization while adapting to the intelligent and dynamic requirements of IoV. The proposed enhanced VCG auction mechanism not only protects the privacy of VSPs but also reduces the likelihood of collusion among VSPs, and it is theoretically proven to be dominant-strategy incentive compatible (DSIC). The simulation results demonstrate that, compared to the VCG mechanism implemented using quantization methods, the proposed mechanism exhibits significant advantages in convergence speed, social welfare maximization, and resistance to collusion, providing new insights into resource allocation in intelligent 6G networks.

\end{abstract}


\IEEEpeerreviewmaketitle 

\section{Introduction}

\IEEEPARstart{W}{ith} the ongoing research and anticipated advancements in 6G network technologies, the Internet of Vehicles (IoV), as one of its core application scenarios, is expected to evolve toward intelligence, high reliability, and low latency \cite{9509294}. 6G networks not only support transmissions with low latency and high data rate but also enable intelligent management of network resources through artificial intelligence technologies \cite{8869705}. In this context, vehicle service providers (VSPs) can offer personalized and high-quality services to users associated with vehicles. In addition to currently available services such as navigation, traffic information prompts, and multimedia services, future services like intelligent driving, driving behavior analysis, and vehicle condition monitoring will significantly enhance traffic efficiency and the driving experience \cite{9088328}. VSPs rely on communication, computation, and cache resources to provide these services. Network service providers (NSPs), with their extensively distributed physical infrastructure, are capable of delivering reliable resource support to numerous VSPs. Nevertheless, the fixed device deployment of NSPs leads to an inflexible network architecture that makes it difficult to adapt to dynamically changing service demands. Network Function Virtualization (NFV) technology can create logically isolated slices dedicated to VSPs based on their specific requirements, significantly enhancing resource utilization while increasing the flexibility of network resource allocation \cite{8907851}. This capability allows the network to dynamically adapt to diverse service types and rapidly respond to evolving vehicular user requirements.

Despite the advantages of NFV, establishing adaptive resource slices for VSPs remains a challenging issue. 
One of the fundamental dilemmas pertains to the competitive dynamics for resources among various VSPs, necessitating that NSPs devise equitable and efficacious mechanisms for resource allocation to foster the active participation of VSPs in resource negotiation. 
Furthermore, given that NSPs often embody public benefit attributes, their optimization goals transcend mere equity in resource distribution, incorporating the maximization of social welfare (the sum of all VSPs' utility). However, the pursuit of societal welfare enhancement may clash with the individual profit-maximization goals of VSPs, thereby adding layers of complexity to the allocation quandary. Such problems are typically modeled within the frameworks of game theory, and auction-based mechanism design emerges as a potential resolution, celebrated for its merits in incentive compatibility and operational efficiency.
When NSPs understand all VSPs’ utility functions, the social welfare maximization problem can be addressed by a classic auction mechanism known as the Vickrey-Clarke-Groves (VCG) auction \cite{MAKOWSKI1987244}. VCG auctions are widely used to achieve efficient allocation in multi-item or multi-attribute auctions, incentivizing bidders to reveal their true valuations for the auctioned items truthfully. In the VCG auction, bidders submit their maximum willingness to pay for each item, and the mechanism selects an allocation that maximizes social welfare. After determining the allocation, the actual payment prices are calculated based on each bidder's impact on the welfare of other bidders.


However, one limitation of existing works on VCG auctions is that calculating the actual payment prices typically requires the utility functions of the bidders \cite{doi:10.1287/opre.1070.0384, 8486341}. 
Specifically,
The utility functions of VSPs in IoV often involve private information that can affect the safety of vehicle operation \cite{9738808, 9360666}, making it difficult for VSPs to report their utility functions to NSPs. 
In addition, utility functions may not have explicit expressions for some complex services. Even if all the utility functions of VSPs possess desirable properties (e.g., convexity), the computational complexity can be significant in large-scale IoV \cite{8486341}. The difficulty increases when utility functions cannot be explicitly expressed. 
Another concern arises from the possibility of collusion among VSPs to maximize their benefits, which makes it challenging for auctioneers to achieve social welfare maximization goals \cite{10.1007/978-3-642-17572-5_4, 10.1145/1160633.1160729}.

To address the issue of utility functions that cannot be explicitly expressed, recent studies have introduced the concept of ``Value of Information (VoI)," which reflects the impact of specific information on the performance of specific services \cite{9541011, SONG2022108034, 9751706}. 
VoI can serve as a transactional medium between NSPs and VSPs, exhibiting currency-like properties for resource allocation purposes.
Compared to directly estimating fixed utility functions, estimating VoI based on current information characteristics can reflect the real-time service performance of VSPs, thereby facilitating the goal of maximizing the performance of all VSP services, i.e., social welfare maximization. 
VoI can support measuring service value without relying on explicit utility functions, facilitating the expression of utility functions.

To address the privacy and collusion issues faced by the auction mechanism, a promising approach is to integrate multi-agent reinforcement learning (MARL) with the auction mechanism. 
MARL can directly learn optimal strategies through environmental interactions without requiring mathematical modeling of the environment \cite{Zhang2021,4445757}.
The MARL-based approach is conducive to VSPs' privacy needs, as NSPs can indirectly understand the relationship between VSPs’ value assessments and resource needs by training neural networks rather than directly acquiring utility functions or data.
The gaming process within the MARL framework introduces randomness \footnote{Randomness in this context is introduced by the exploration process of MARL, meaning that the allocation schemes obtained in the same state exhibit a certain level of uncertainty. This characteristic makes it more challenging for conspirators to achieve a stable allocation strategy (reaching an equilibrium point in the game).} into the auction mechanism, increasing the difficulty of bidder collusion and reducing its impact \cite{https://doi.org/10.1111/deci.12159, CHE2009565, HU201184}.

This paper investigates the problem of establishing VSP resource slices in IoV and proposes an improved VCG auction mechanism based on mean-field multi-agent reinforcement learning (MFMARL). 
Unlike the usual VCG auction mechanism, a third-party institution, known as the ``Bank", is introduced to issue VoI Currency (VC), impose taxes, and reclaim VC. 
The mechanism can be divided into three phases: currency issuance, sealed bidding for intention value, and information disclosure. 
During the currency issuance phase, the Bank determines the amount of VC to issue based on the total resources of the NSP and historical social welfare, establishing a reserve pool. 
In the sealed bidding phase, VSPs submit their resource bidding intentions in VC to the NSP and borrow VC from the Bank based on their expectations. In the information disclosure phase, the NSP determines the resource allocation scheme based on the bidding intentions and discloses the information, after which the VSPs pay VC to acquire resources and begin providing services. Finally, the Bank calculates taxes and reclaims residual VC based on the information disclosed by the NSP and the financial statements of the VSPs. 
Throughout the process, the NSP and the Bank are training their strategies within the MFMARL framework to achieve near-optimal social welfare allocation while protecting VSP privacy and reducing the impact of collusion. The main contributions of this paper are as follows:


\begin{itemize}
\item An improved VCG auction mechanism based on MFMARL was designed, achieving near socially optimal resource allocation without requiring knowledge of VSP utility functions. 
Additionally, the mechanism can be scaled to changes in NSP available resources and the number of VSPs without the need for redeployment.

\item 


Theoretical analyses and simulation studies assert that the proposed auction mechanism introduced achieves Dominant Strategy Incentive Compatibility (DSIC), indicating that for all bidders(VSP), strategies that optimize the social welfare equally maximize their profits. This secures the initiative of the VSPs to participate in the auction process.

\item 

Theoretical analyses and simulation studies confirm the effectiveness of the proposed auction mechanism in addressing the collusion of bidders, reducing the negative impact of such collusion on overall social welfare, and showing enhanced robustness. 

\end{itemize}

The rest structure of this paper is organized as follows: Section II reviews and summarizes the previous work on VCG auctions and MRAL-enhanced auctions. In Section III, we provide a comprehensive description of the system model. Section IV introduced the proposed auction mechanism where information value serves as currency and analyzes its theoretical performance. Part V elucidates the implementation method of the auction mechanism based on MFMARL. Section VI demonstrates the results of simulations, and Section VII concludes the paper.

\section{Related work}

NFV brings numerous benefits to IoV, including isolation, automation, flexibility, and scalability. 
Many researchers have attempted to work in this research field from economics and game theory perspectives. 
In \cite{5606180}, a dynamic spectrum leasing mechanism for power control strategies was proposed, while a spectrum-sharing power allocation scheme based on a Stackelberg game for Femtocell networks, regulated by pricing, was introduced in \cite{6171995}. Moreover, utility maximization in spectrum-sharing networks using game-theoretic power control strategies was explored in \cite{4570231} and \cite{4674354}. A scenario in which mobile users compete for wireless resources, aiming to optimize profits for both physical substrates and virtual wireless networks, was studied in \cite{7086818}.
Although these game-theoretic approaches drive resource competition towards equilibria such as Nash equilibrium, they often fail to achieve social welfare maximization.
Consequently, researchers have further refined auction mechanisms.

\subsection{VCG-based auction mechanism}

The VCG auction, a non-cooperative game strategy, is widely recognized for its ability to maximize social welfare. Its application to network slice resource allocation has been extensively studied. For example, \cite{7504529} proposed a model based on VCG auctions, where NSPs act as auctioneers to allocate discrete physical resource blocks. Similarly, \cite{7127601} integrated a VCG auction model to address power allocation in LTE air interface virtualization. However, these approaches often face challenges due to limitations in information exchange mechanisms and the non-convexity of bidders' utility functions, hindering the practical implementation of VCG auctions.


To address these issues, some scholars have focused on allocating infinitely divisible resources under limited information exchange mechanisms. For instance, \cite{https://doi.org/10.1002/ett.4460080106} and \cite{doi:10.1057/palgrave.jors.2600523} proposed Kelly's proportional fairness mechanism, where agents submit one-dimensional bids, and allocations are determined based on the ratio of bids to a congestion price. However, this approach may incur efficiency losses of up to $25\%$ of the maximum social welfare, as noted by \cite{doi:10.1287/moor.1040.0091}. To mitigate this drawback, several works have combined Kelly's concept with the VCG auction, resulting in the ``VCG-Kelly" mechanism \cite{4278423, ma2010resource, doi:10.1287/opre.1080.0638}. The ``VCG-Kelly" mechanism requires agents to communicate only a single scalar value, enabling the allocation of infinitely divisible resources under limited information exchange. Furthermore, \cite{8486341} proposed a dominant allocation strategy for divisible network resources under constrained information by partitioning resources into indivisible units and applying the VCG mechanism to each partition, achieving near-socially optimal allocations.

\subsection{MARL-based auction mechanism}

In recent years, MARL has gained attention for its potential in auction modeling. MARL provides a model-free approach to decision-making and control, enabling agents to learn optimal strategies through interaction with the environment without prior knowledge of the system. For example, \cite{10403539} used MARL to reduce operating costs in double auctions within energy markets, while \cite{10538064} integrated the Multi-Agent Deep Deterministic Policy Gradient (MADDPG) algorithm with electricity market auction mechanisms to guide participants toward better bidding strategies. However, many MARL algorithms suffer from poor scalability, limiting their applicability to vehicular network resource slicing problems with rapidly changing environments. Although independent reinforcement learning approaches can mitigate this issue, they often exacerbate non-stationarity, hindering stable strategy development.


To address these challenges, extended MADDPG frameworks have been proposed. For instance, \cite{77884806b6ff45a88f2711365e3d698d} introduced an order book structure that allows agents to abstract and observe the behaviors of others. However, implementing such frameworks becomes increasingly difficult as the number of agents grows. The emergence of MFMARL offers a promising solution, providing robust scalability and reducing computational requirements \cite{pmlr-v80-yang18d}. In IoV, MFMARL has been successfully applied to address joint spectrum and power allocation problems, significantly reducing computational complexity \cite{9919273}.

Despite these advancements, the adoption of the aforementioned two types of auction mechanisms still encounters certain limitations. Auction mechanisms based on the VCG principle often require substantial communication to comprehend the utility functions of participants, posing significant demands on the communication capabilities of IoV networks and failing to meet the privacy requirements of VSPs; concurrently, auction mechanisms based on MARL typically face limitations in scalability.
This paper builds upon and extends these ideas, aiming to provide innovative solutions to the resource allocation challenges in vehicular network slicing.

\section{System Model and Problem Formulation}

\begin{figure}[htpb]
    \centering
    \includegraphics[width=3.4in]{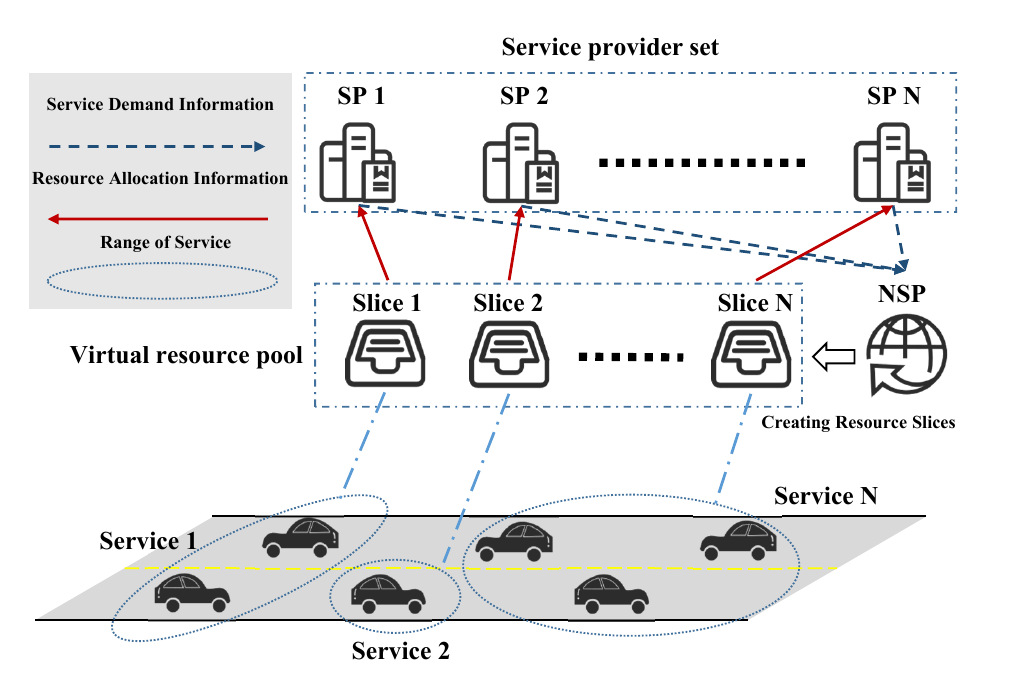}
    \caption{Schematic illustration of how IoV resources are sliced and allocated.}
    \label{fig:system_model}
\end{figure}

We consider an IoV composed of one NSP and \( N \) VSPs as shown in Fig. \ref{fig:system_model}. Base stations (BS), mobile edge computing (MEC) servers, and roadside units (RSU), are owned by the NSP. The communication, computation, and cache resources are virtualized into a resource pool of the NSP.
These virtualized network resources can be partitioned into slices by the NSP and allocated to VSPs for a certain period. 
Each period, denoted as time \(T_{s}\), is characterized by VSPs submitting their bids to the NSP, which in turn allocates the new resource slices to the VSPs. The VSPs aim to maximize profits, whereas the NSP allocates resources to maximize social welfare. 

\subsection{VSP Profits Model}
We denote the VSP set as \( \bm{\mathcal{V}} \), and define VSP \( i \) as \( v_{i} \). The slice set is denoted by \( \bm{\mathcal{S}}\). The slice vector in $t$-th round is denoted by $\bm{s}_t$, and \( s_{i,t} \) denotes the resource slice allocated by the NSP to \( v_{i} \) in $t$-th round, constituted by bandwidth resources, computational resources, and cache resources. 
The terms $\bm{{X}}$, $\bm{x}_t$, \( x_{i,t} \) signify the bid set, the bid vector in $t$-th round and the bid of the VSP $i$ in $t$-th round. 
For VSP \( i \), we employ \( P_{i,t} \) to represent the profit of VSP \( i \) during the \( t \)-th round, with \( P_{i,t} \) being calculated as
\begin{align}
    \label{eq:proft}
    P_{i,t}(x_{i,t}) = g_{i}(s_{i,t}) - z_{i}(x_{i,t}),
\end{align}
    where \( g_{i}(\cdot) \) serves as the gross profit function. Note is that the gross profit function encompasses the private information of the VSP and may not always offer a manifest expression. The term \( z_{i}(\cdot) \) represents the cost function of deploying the resource slice based on the bid.

\subsection{Social Welfare Model}
The maximization of social welfare symbolizes the maximization of the aggregate utility of all VSPs. 
Momentous to note is the discrepancy between utility and profits\footnote{The discussion of the utility notion bears unique significance in IoV as several services within these networks pertain to vehicular driving safety, such as autonomous driving, assisted driving, navigation, etc. These services may not generate substantial profits but proffer immense utility (i.e., they may prevent severe traffic accidents).}, 
implying that an accurate utility function is generally unobtainable directly through the gross profit function (even if the profit function is explicative).
We let \( U_{i,t} \) to denote the utility of VSP \( i \) in the \( t \)-th round and employ the function \( u_{i}(\cdot) \) to express the bond between VSP \( i \)'s utility and the resource slice. For the NSP, we define the total social welfare $\mathcal{SW}_t$ as 
\begin{align}
    \mathcal{SW}_t = \sum_{i = 1}^{N} U_{i,t} = \sum_{i = 1}^{N}u_{i}(s_{i,t}).
\end{align}
Analogous to the gross profit function, the utility function may likewise be implicit; a more dire circumstance is that neither the NSP nor the VSPs might understand the specific form of the utility function\footnote{This owes to the more subjective nature of utility compared to profits, usually involving insuperably abstract mapping measures such as user satisfaction.}. 

Normalization of the resources within the resource pool is performed for convenient analysis, ensuring that \( \sum_{i = 1}^{N}{s_{i,t}} \leq \mathds{1} \). Although the precise form of the utility function is arduous to obtain via the gross profit function, an estimate can still be fabricated through the observation of gross profits. Hence, we establish the mapping from the utility function to the gross profit function as
\begin{align}
    \label{gross p}
     G_{i,t} = \alpha_{i}U_{i, t}  + {w}_{i}, 
\end{align}
where \( \alpha_{i} \) is the rate coefficient of gross profit $G_{i,t}$ to utility $U_{i, t}$, possessing varied values in different VSPs, and \({w}_{i} \) is a random variable obeying a specific distribution representing the uncertain association between profit and utility. Consequently, we can rewrite the profit function as
\begin{align}
     P_{i,t}(x_{i,t}) = \alpha_{i}u_{i}(s_{i,t}) + {w}_{i} - z_{i}(x_{i,t}).
\end{align}

To impel the achievement of maximal social welfare, it is requisite that a consensus is conceivably reached between VSP and NSP. In simpler terms, the socially optimal allocation result must remain dominant \footnote{In game theory, a dominant strategy refers to a course of action that yields the optimal outcome for a participant, irrespective of the strategies employed by other players. In other words, selecting a dominant strategy is invariably the most advantageous move, regardless of the opponents' actions.} for both VSPs and the NSP. This consensus is represented mathematically as 
\begin{align}
     \bm{s}^{\ast}_t = \mathcal{A} (x^{\ast}_{1,t}, \ldots,x^{\ast}_{i,t}, \ldots).
\end{align}
Due to both gross margin functions and utility functions being unknown to the NSP, we utilize function $\mathcal{A}(\cdot)$ to represent an abstracted process of resources' allocation by NSP based on the bid from VSPs. 
\(x^{\ast}_{i,t} \) denotes the optimal bid under the circumstance of VSP maximizing its profits as
\begin{align}
     x^{\ast}_{i,t} = \arg\max P_{i,t}(x_{i,t}),
\end{align}
while \( \bm{s}^{\ast}_{t} \) signifies the optimal resource slice allocation under the condition of maximal social welfare in $t$-th round as
\begin{align}
     \bm{s}^{\ast}_{t} = \underset{\bm{s}_{t}}{\arg\max} \sum_{i = 1}^{N} u_{i}(s_{i,t}).
\end{align}

\subsection{Problem Formulation}
We formalize the problem of resource slicing in IoV as
\begin{problem}
    \begin{align}
        \label{P1}
        \mathcal{P}1: &~~ \underset{\bm{\mathcal{S}}}{\max}\sum_{t = 1}^{T}\sum_{i = 1}^{N} u_{i}(s_{i,t}), \\
        &s.t. ~ \sum_{i = 1}^{N}{s_{i,t}} \leq \mathds{1}, \tag{\ref{P1}a} \label{C1}\\
        &~~~~ \bm{s}_{t} = \mathcal{A} (x^{\ast}_{1,t}, \ldots,x^{\ast}_{i,t}, \ldots), \tag{\ref{P1}b} \label{C2}\\
        &~~~~ x^{\ast}_{i,t} = \arg\max P_{i,t}(x_{i,t}).\tag{\ref{P1}c} \label{C3}
    \end{align}
\end{problem}

Constraint (\ref{C1}) ensures that the slices do not exceed the available resources of the NSP, while (\ref{C2}) and (\ref{C3}) represent the strategic consensus that can be reached between the NSP and the VSP. 
The intricacy of resolving problem \(\mathcal{P}1\) stems from the NSP inability to directly obtain formulations for utility or profit functions. This limitation affects the development of effective slicing allocation strategies.
Fortunately, the NSP can deduce the relationship between resource slicing, social welfare, and VSP profit through bid information from the VSP and statistical information on overall utility over a while. This offers us a viable avenue for addressing the problem above.

\section{An Enhanced Dual-Currency VCG Auction Mechanism Based on VoI}

We propose a dual-currency auction mechanism based on the VoI, an extension of the traditional VCG auction. In this section, we first elucidate the implementation of this mechanism and subsequently demonstrate that it shares a similar DSIC with VCG auctions.
\subsection{Auction Mechanism}
\begin{figure*}[htbp]
    \centering
    \includegraphics[width=7.0in]{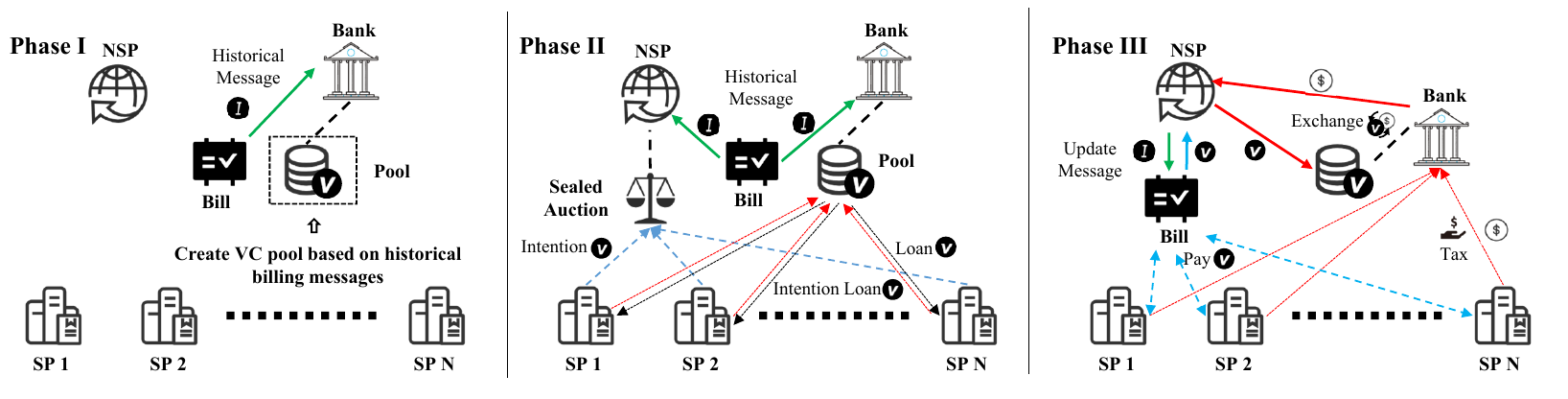}
    \caption{Three phases of VCG auction mechanism for dual currency.}
    \label{fig:auction_mesh}
\end{figure*}

Firstly, we explain the concept of the VoI. As previously discussed, one of the challenges of problem \(\mathcal{P}1\) lies in the uncertainty of the utility function, which precludes an accurate mapping to profit. The VoI metric, as described in \cite{9541011}, is defined based on the impact of specific information on the performance of a particular task, which aptly reflects the statistical information of task performance (i.e., utility in this context). 
We extend this concept in this paper, where VoI represents the expected utility of VSP under the social welfare maximization strategy $\pi$. The specific formulation is as
\begin{align}
    \text{VoI}_{i,t} &= \mathbf{E}\left[U_{i,t} | \pi \simeq \max \sum_{i = 1}^{N} u_{i}(s_{i,t})  \right].
\end{align}

By understanding the VoI distribution of the VSP, the NSP can indirectly comprehend its utility function. To this end, we design a dual-currency auction mechanism based on VoI, as illustrated in Fig. \ref{fig:auction_mesh}. Unlike traditional auction mechanisms, we introduce a third-party entity called the ``bank". This bank can be stationed in neutral regulatory department devices or operate on the NSP's virtual machine. Irrespective of its deployment location, the bank must ensure transparency in its information (bill), implying that all VSPs can scrutinize it directly. In tandem, the bank collects publicly available information $\mathcal{B}_{t} = (U_{t}, G_{t})$
from VSPs' gross profits and feedback on user satisfaction, which we term ``financial reports". 
It is important to note that this information exhibits a delay, which means that in the \(t\)-th round of the auction, only the information from the previous \(t-1\) rounds can be observed.
We presume the bank's historical bills, constituting the prior information, are represented as $\mathcal{BI}$. The specifics of the auction mechanism can thus be articulated as follows:
\begin{itemize}
    \item During the currency issuance phase, the bank estimates the mean total VoI of all VSPs based on historical bill information $\mathcal{BI}$. It forms a pool $\mathcal{O}(t)$ of VoI currency (VC) as
    \begin{align}
      \mathcal{O}(t) &= \sum_{i = 1}^{N} \mathbf{E}\left[U_{i,t} | \pi \simeq \max \sum_{i = 1}^{N} u_{i}(s_{i,t}) \right] \\ \notag
      & = \sum_{i = 1}^{N} f_{i}(\mathcal{BI};\hat{\phi}_{i}),
    \end{align}
    where the function \(f_{i}(\cdot)\) signifies the bank's projection of the utility achievable by the VSP $i$ under the stratagem of optimizing its profit, with \(\hat{\phi}_{i}\) representing the inferred of the function parameters.
    
    \item In the sealed bid for intention value phase, each VSP sends proposals $x_{i,t}$ to the NSP in units of VC, reflective of their service profit estimates, and applies for a loan of equal VC amount from the bank. The bank allows loans $\mathcal{K}_{t}$ to VSPs based on the $\mathcal{O}(t)$. $\mathcal{K}_{t}$ can be calculated as
    \begin{align}
     \mathcal{K}_{t} = (\mathcal{K}_{1,t},\cdots,\mathcal{K}_{N,t}) = \mathcal{O}(t) \ast \frac{x_{i,t}}{\sum_{j = 1 }^{N}\bm{x}_{j,t}}.
    \end{align}

    \item The NSP allocates slices and estimates the VSP's utility based on the submitted proposals and historical account bills $\mathcal{BI}$. The method for achieving the optimal slicing policy and optimal sum utility $\mathcal{SW}^{\ast}_{t}$ in the $t$-th round is 
    \begin{align}
    \label{P2}
        (\mathcal{SW}^{\ast}_{t}, \bm{s}^{\ast}_{t}) = \underset{\bm{s}_{t}}{\arg\max}\sum_{i = 1}^{N}{u}_{i}(s_{i,t};\hat{\theta}_{i}) ,
    \end{align}
    \begin{align}
    \label{allocate}
       \bm{\tilde{s}}_{t}^{} = \text{norm}\left(\bm{s}^{\ast}_{t} +  \mathcal{N}(0, \Delta^2 \bm{I})\right),
    \end{align}
    \begin{align}
        \Delta =\frac{\text{max}\left(\sum_{i = 1}^{N}x_{i,t} - \mathcal{SW}^{\ast}_{t}, 0\right)}{N}.
    \end{align}
    Where \(\theta_{i}\) represents the estimated parameters of the function \(u_i(\cdot)\), \(\Delta\) represents the amount of additional resources to be allocated to each VSP in instances where the aggregated bids from all VSPs exceed the maximum social welfare. If the aggregated bids do not exceed this threshold, \(\Delta = 0\). The expression \(\mathcal{N}(0, \Delta^2 \bm{I})\) denotes a multivariate Gaussian random variable with a mean of zero and a variance of \(\Delta^2\), utilized to add stochastic elements to each allocation of resources. The function \(\text{norm}(\cdot)\) ensures that the total of the resource allocation vector is kept within the permitted bounds, hence adhering to the established resource constraints.
    
    \item In the phase of information disclosure, the NSP computes the VC interest that VSP \(i\) is required to pay during round \(t\) as \(|\mathcal{K}_{i,t} - U_{i,t}^{\ast}|\), where \(U_{i,t}^{\ast} = u_i({s}^{\ast}_{i,t}; \hat{\theta}_i)\).
    In addition, the social welfare impact of VSP $i$ on other participants can be calculated as
    \begin{align}
    \label{impact}
        \mathcal{I}_{i,t} = & \left(\underset{\bm{s}_{t}}{\max}\sum_{j = 1, j \neq i}^{N}{{u}_{j}(s_{j,t};\hat{\theta}_{j})}\right) 
        -  \sum_{j = 1, j \neq i}^{N}{{u}_{j}(\tilde{s}_{j,t};\hat{\theta}_{j})}.
    \end{align}

    \item Subsequently, the NSP amalgamates the VC interests of the VSP with their repercussions on societal welfare, imposing a tangible financial tax \( Z_i \) on the VSP. The methodology for computing this taxation is delineated as 
    \begin{align}
    \label{tax}
        Z_{i,t} = \hat{\alpha}_{i}(\mathcal{I}_{i,t}  + |\mathcal{K}_{i,t} - U_{i,t}^{\ast}|).
    \end{align}

    \item Finally, the bank will accumulate all openly accessible bill information from the current auctioning round, appending it to pre-existing bills denoted as \(\mathcal{BI}^{+}\) while reevaluating all parameters. The specifics of the collected information and updated parameters include the following parts
    \begin{align}
        &\mathcal{BI}^{+} = \mathcal{BI} \cup (\bm{\tilde{s}}_{t}, \bm{x}_{t}, \mathcal{K}_{t}, U^{\ast}_{t}, \mathcal{I}_{t}, \mathcal{B}_{t}), \\
        &(\hat{\bm{\phi}}, \hat{\bm{\alpha}}, \hat{\bm{\theta}}) \to \underset{\bm{\phi}, \bm{\alpha}, \bm{\theta}}{\arg\min}\mathcal{L}(\bm{\phi}, \bm{\alpha}, \bm{\theta}, \mathcal{BI}^{+}) ,
    \end{align}
    where parameters $\bm{\phi}, \bm{\alpha}, \bm{\theta}$ exemplify the parameter vectors constituted by corresponding parameters, while $\hat{\bm{\phi}}, \hat{\bm{\alpha}}, \hat{\bm{\theta}}$ denote the inference pertaining to the aforementioned parameter vector. The function $\mathcal{L}(\cdot)$ is the loss function used for parameter estimation.
\end{itemize}

\subsection{Dominant Strategy Incentive Compatible Analysis}

DSIC, an essential concept originating from game theory and mechanism design, entails a circumstance in which the architect's goal is to establish an environment wherein participants' optimal actions, or their ``dominant strategies", are in unison with the objectives of the design architect \cite{MAKOWSKI1987244}. In this paper, our design target for this auction mechanism is to optimize social welfare. Hence, an auction mechanism possessing the DSIC attribute implies that the maximization of social welfare aligns with the VSP's aim of profit maximization. 

\begin{figure*}[h]
    \begin{align}
    \label{expect_p}
        \mathbf{E}[P_{i}|x_{i}] &= \alpha_{i} u_{i}(\tilde{s}_{i};\theta) + \mu({w_{i}}) + \mathbf{E}\left[\hat{\alpha}_{i}|x_{i}\right] \left( \sum_{j = 1, j \neq i}^{N}{{u}_{j}(\mathbf{E}\left[\tilde{s}_{j}|x_{i}\right];\mathbf{E}\left[\hat{\theta}_{j}|x_{i}\right])} \right) 
        - \mathbf{E}\left[\hat{\alpha}_{i}|x_{i}\right]\left( \underset{\bm{s}}{\max}\sum_{j = 1, j \neq i}^{N}{{u}_{j}(s_{j};\mathbf{E}\left[\hat{\theta}_{j}|x_{i}\right])}\right) \notag \\ 
        &~~- \mathbf{E}\left[\hat{\alpha}_{i}|x_{i}\right] \left|\sum_{i = 1}^{N} f_{i}(\mathcal{BI};\mathbf{E}\left[\hat{\phi}_{i}|x_{i}\right]) \ast \frac{x_{i}}{\sum_{j = 1 }^{N}\bm{x}_{j}} - u(s_{i}^{\ast};\mathbf{E}\left[\hat{\theta}_{i}|x_{i}\right])\right| \notag \\ 
        &= \underbrace{\mu({w_{i}}) - \alpha_{i}\mathcal{SW}_{-i}^{\ast}}_{\textbf{Item} 1} + \underbrace{\alpha_{i} \left(\mathcal{SW}(\bm{\tilde{s}}) - \left|\sum_{j = 1}^{N} f_{j}(\mathcal{BI};\phi_{j}) \ast \frac{x_{i}}{\sum_{j = 1 }^{N}\bm{x}_{j}} - u_{i}(s_{i}^{\ast};\theta_{i})\right|\right)}_{\textbf{Item} 2}
    \end{align}
    \vspace{2pt}
    \hrulefill
\end{figure*}

As articulated in Section III, our problem essentially maps to a multi-round auction scenario. For an analysis of the DSIC properties of the auction mechanism,  it's beneficial to consider the VSP's expected profit during the $T$-round auction when $T \to \infty$. Here, we operate under the assumption that the mean value of the random variable \(w_{i}\) is designated as \(\mu(w_{i})\), and the estimator of the parameters $(\bm{\phi}, \bm{\alpha}, \bm{\theta})$ are unbiased. The expected profit for a VSP's bid can thus be represented as
\begin{align}
    \mathbf{E}[P_{i}|x_{i}] &= \alpha_{i} u_{i}(\tilde{s}_{i};\theta) + \mu({w_{i}}) \\  \notag
        &- \mathbf{E}[\hat{\alpha}_{i}(|\mathcal{K}_{i} - U^{\ast}_{i}|  + \mathcal{SW}^{\ast}_{-i}  - \mathcal{SW}_{-i}(\bm{\tilde{s}}) ~|x_{i}],
\end{align}
where $\mathcal{SW}^{\ast}_{-i} = \underset{\bm{s}}{\max}\sum_{j = 1, j \neq i}^{N}{{u}_{j}(s_{j};\hat{\theta}_{j})} $ represents optimal social welfare allocation result except VSP $i$, and $\mathcal{SW}_{-i}(\bm{\tilde{s}}) = \sum_{j = 1, j \neq i}^{N}{{u}_{j}(\tilde{s}_{j};\hat{\theta}_{j})}$ represents current social welfare except VSP $i$ .
For brevity in our exposition, we have intentionally omitted the round marker, $t$, from the formula above. 
As the auction rounds converge towards the endless, we can assume that the sample set $\mathcal{BI}$ from historical records assumes a large magnitude. 
This ensures the estimation of parameters $(\bm{\phi}, \bm{\alpha}, \bm{\theta})$ independent of the VSP bidding stratagem because of our auction mechanism. Therefore, the expected profit for VSP can be simplified to the following form as equation (\ref{expect_p}).

Item 1 of the equation (\ref{expect_p}) remains irrelevant to VSP's bid. In term 2, since the parameters $\mathbf{\theta}$ and $\mathbf{\phi}$ are unbiased, as the number of auction rounds tends to infinity, the VoI of VSP $i$ which corresponds to the function $f_{i}(\mathcal{BI};\phi_{i})$ will tend to $u_{i}(s_{i}^{\ast }; \theta_{i})$. Then, the item 2 can be rewritten as

\begin{align}
    \mathcal{SW}(\bm{\tilde{s}}) - \left|\sum_{j = 1}^{N} u_{j}(s_{j}^{\ast};\theta_{i}) \ast \frac{x_{i}}{\sum_{j = 1 }^{N}{x}_{j}} - u_{i}(s_{i}^{\ast};\theta_{i})\right|,
\end{align}
and we discern item 2 under two cases. 

\textbf{Case I}: Given a VSP $i$, when other VSPs disclose their genuine VoI as the bid. item 2 can be rewritten as
\begin{align}
    \mathcal{SW}(\bm{\tilde{s}}) - \left| \frac{x_{i} \ast \sum_{j = 1}^{N} u_{j}(s_{j}^{\ast};\theta_{i})}{x_{i} + \sum_{j = 1, j\neq i }^{N}u_{j}(s_{j}^{\ast};\theta_{i})} - u_{i}(s_{i}^{\ast};\theta_{i})\right|.
\end{align}
The dominant strategy for VSP \(i\) lies in bidding equivalent to its authentic value of VoI. In this case, the SW depends on the bid \(x_i\) of VSP. When VSP \(i\)'s bid exceeds the true VoI, $\mathcal{SW}(\bm{\tilde{s}})$ is admittedly less than the optimum \(\mathcal{SW}^{\ast}\), and when \(x_i\) is lower than the true VoI, $\mathcal{SW}(\bm{\tilde{s}})$ is equivalent to \(\mathcal{SW}^{\ast}\). Due to the presence of the absolute value terms, \(x_{i} = u_{i}(s_{i}^{\ast}; \theta_{i})\) is the dominant strategy.

\textbf{Case II}: When some VSPs abstain from tendering their actual VoI as a bid, it is construed that a state of collusion exists amongst the VSPs. Without loss of universality, we postulate that the assembly of VSPs participating in such collusion is denominated as set $\mathcal{M}$. 
It's essential to underscore that for the $\mathcal{SW}(\bm{\tilde{s}})$, even under collusive conditions, the VSP has no incentive to elevate the bid beyond the true VoI. Because under parameter-unbiased circumstances, the $\mathcal{SW}(\bm{\tilde{s}})$ cannot transcend the optimum \(\mathcal{SW}^{\ast}\). 
Under such conditions, the absolute value part for any given VSP $i$ in $\mathcal{M}$ can be expressed as
\begin{align}
    \left| \frac{x_{i} \ast \sum_{j = 1}^{N} u_{j}(s_{j}^{\ast};\theta_{i})}{x_{i} +\sum_{m \in \mathcal{M}, m\neq i} x_{m} + \sum_{j = 1, j \notin \mathcal{M} }^{N}u_{j}(s_{j}^{\ast};\theta_{i})} - u_{i}(s_{i}^{\ast};\theta_{i})\right|.
\end{align}
Drawing upon the characteristics of absolute values, it becomes apparent that the lowest point of the item occurs at \(0\), under which conditions \(x_i\) is determined as
\begin{align}
    \label{collusion_x}
    x_{i} = \frac{u_{i}(s_{i}^{\ast};\theta_{i})\left(\sum_{m \in \mathcal{M}, m\neq i} x_{m} + \sum_{j = 1, j \notin \mathcal{M} }^{N}u_{j}(s_{j}^{\ast};\theta_{i}) \right)}{\sum_{j = 1, j\neq i}^{N} u_{j}(s_{j}^{\ast};\theta_{i})}
\end{align}



Within the set \(\mathcal{M}\), for VSPs to achieve collusion, a unified depreciation of the item by all members is essential. Without such depreciation, any VSP that fails to reduce the item to its minimal value finds no incentive to engage in collusive activities. Consequently, the bidding behavior of each VSP in set \(\mathcal{M}\) must conform to equation (\ref{collusion_x}). This stipulation leads to a system of linear equations characterized by \(\mathcal{M}\) variables, expressed as \(\mathbf{A} x_{\mathcal{M}} = \mathbf{b}\), where \(\mathbf{A}\) is a matrix of full rank. This system, therefore, yields a unique solution.

From Case I, we observe that \(x_i = u_i(s_i^*; \theta_i)\) presents itself as a non-zero solution within the system. Together with the specified conditions, it also stands as the sole non-zero solution. Therefore, it can be concluded that in scenarios where VSPs collude, tendering their genuine VoI in bids retains its status as a dominant strategy.

\section{MARL Algorithm}

To enact the mechanism articulated in Section IV, a cardinal issue is estimating parameters and resolving the optimization problem delineated in (\ref{P2}). The parameter \(\alpha_{i}\) can be estimated unbiasedly using the least squares method. 
The parameters \(\bm{\theta},\bm{\phi}\) and the optimization problem (\ref{P2}) can be interpreted as a variant of the multi-armed bandit problem. 
The VSPs can be seen as several arms, where the payoff of each arm corresponds to the societal welfare provided in any given slice allocation scenario. Thus, the problem can be modeled as a Markov Decision Process (MDP) problem, wherein its states, actions, and rewards are defined as follows:

\textbf{State}:
We designate the resource slice allocation \(\bm{s}_{t-1}\)
as the observation state \(\bm{o}_{t}\) for round \(t\). Hence, the $\bm{s}_{t-1}$ consists of a \(3 \times N\) matrix, where \(N\) stands for the collective total of VSPs, and \(3\) denotes the varieties of resources.

\textbf{Action}:
We stipulate the action as the decision on adjusting resource allocation based on the current observation state \(\bm{o}_{t}\), i.e., \(\bm{a}_{t}\) - also a \(3 \times N\) matrix. Each element of this matrix has a discrete value in the $[-0.3, -0.2, -0.1, 0, 0.1, 0.2, 0.3]$, illustrating the proportion of the corresponding VSP's resource allocation either increased or decreased when juxtaposed with the last round. To adhere to constraint (\ref{C1}), a normalization map is necessitated on the outcome of the adjusted resource allocation.

\textbf{Reward}:
After a round of slicing segmentation, the reward of the MDP is the cumulative bid sum gained by all VSPs as 
\begin{align}
    \mathcal{R}_{t} = \sum_{i = 1}^{N} x_{i,t},
\end{align}

Consequently, we are capable of employing RL methodologies to train a network, designed for the optimization of resource slicing partitioning, to stand instead of equation (\ref{P2}) and parameter \(\bm{\theta}\). However, in the design phase of said reinforcement learning strategies, we are confronted with a pair of issues. The first pertains to scalability; as previously mentioned in Section I, the quantity of VSPs is subject to frequent modification. We aspire to craft an auction mechanism that adaptively accommodates the variability in the number of VSPs. For conventional neural networks (such as MLPs), this implies transformative changes to the network's width, necessitating a process of retraining and redeployment. 
The second predicament is found within equation (\ref{impact}), in which there is a requisite to obtain the socially optimal allocation devoid of VSP $i$. This, by implication, necessitates not a unified neural network, but rather \(N\) individual networks that are amenable to arbitrary combinations.
On the other hand, due to the actions of the MDP being enshrined within a $3\times N$ matrix, this insinuates the possibility of the action space becoming exceedingly vast, particularly in circumstances wherein the count of VSPs is considerable, thus giving rise to the dimensional curse amid learning and optimization procedures.

To address the above complexities, the MFMARL framework permits the training of $N$ neural networks as agents, and employing these agents to infer socially optimal allocation schemes of any combination. For convenience, we still use the VSP numbers to label agents. Concurrently, with the aid of mean field theory, we can substantially mitigate the dimensionality of the joint action space. Within MFMARL, the joint action-value function of agent $i$ can be deconstructed as follows:
\begin{align}
\label{Q_function}
    \mathcal{Q}_{i}(\bm{o}_{t}, \bm{a}_{t}) &= \frac{1}{\mathcal{N}^{i}} \sum_{j \in \mathcal{N}(i)}\mathcal{Q}_{i}(o_{i,t}, a_{i,t}, m_{j,t}) \notag \\
    &
    \approx Q_{i}(o_{i,t}, a_{i,t}, \overline{m}_{i,t}) 
\end{align}
where \(\overline{m}_{i,t} = \frac{1}{\mathcal{N}^i} \sum_{j \in \mathcal{N}(i)} m_{j,t}\),  $m_{j,t} = \overline{m}_{i,t} + \delta m^{i,j}$ and $m_{j,t} = (o_{j,t}, a_{j,t})$ represents the one-hot encoding of agent \(j\)'s discrete action $a_{j,t}$ and state $o_{i,t}$, that is, if $a_{j,t} = -0.3$ \(\overline{m}_{j,t} \triangleq [1,0,0,0,0,0,0, o_{i,t}]\). 
$\delta m^{i,j}$ is a small fluctuation between $m_{j}$ and mean-field \(\overline{m}_{i}\).
$\mathcal{N}(i)$ denotes the neighborhood set of agent $i$.
\(\overline{m}_{i}\) can be interpreted as the empirical distribution of the state and the actions taken by agent \(j\)'s neighbors, equivalently, an estimation of the action distribution within the cohort of agents in the mean-field. The proof for equation (\ref{Q_function}) is detailedly expounded in \cite{pmlr-v80-yang18d}.

\begin{figure*}[htbp]
    \centering
    \includegraphics[width=7.0in]{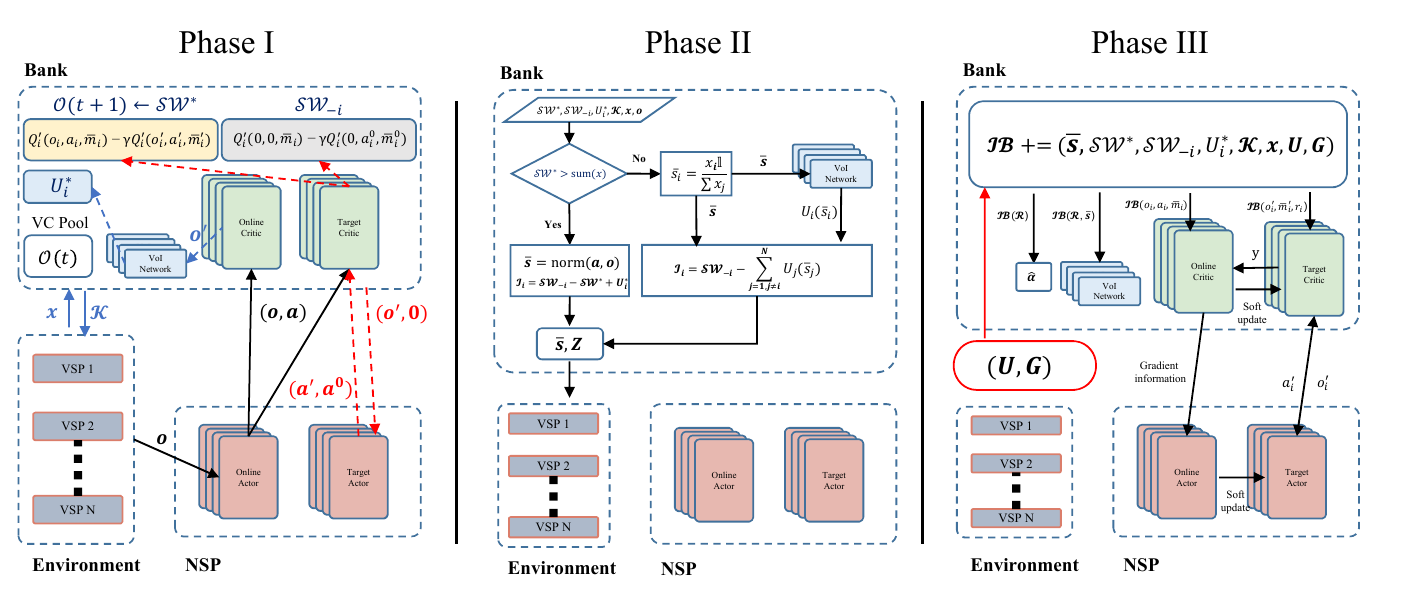}
    \caption{Three phases of VCG auction mechanism for dual currency under MFMARL framework.}
    \label{fig:train}
\end{figure*}
The agents' networks employ a dual Actor-Critic architecture, with each agent possessing a target and online network. 
Both the online actor and target actor are deployed at the decision center of the NSP, whereas the online critic and the target critic are positioned within the bank. The auction mechanism under the MFMARL framework manifests as illustrated in Fig. \ref{fig:train}.
Consistent with our descriptions in Section IV, the entire auction process is divided into three distinct phases. In the initial phase, the VSP dispatches a bid \( \bm{x}_{t} \) to the bank and secures the loan \( \mathcal{K}_{t} \). Meanwhile, the NSP, guided by the observed state \( \bm{o}_{t} \), employs the online actor to acquire an action \( \bm{a}_{t} \). We employ \( \mu_{i}(o_{i,t}, \overline{o}_{i,t}) \) and \( \mu_{i}^{'}(o_{i,t}, \overline{o}_{i,t}) \) to represent the online actor and target actor networks respectively, the input to the actor-network being the observed state \( o_{i} \) to wit, the resource slices apportioned to VSP \( i \) during previous rounds, and the average slices allocated to each VSP excluding \( i \), notated as \( \overline{o}_{i,t} \).

To compute the taxes payable by VSP \( i \), the NSP presents the joint observed states \( \bm{o}_{t} \) and actions \( \bm{a}_{t} \) to the bank. The bank, in response, employs the critic network nested within it to estimate the optimal societal welfare  \( \mathcal{SW}_{-i} \) without VSP \( i \). Corresponding to the actor network, the critic network also encompasses online and target versions, signified as \( Q_{i}(o_{i,t}, a_{i,t}, \overline{m}_{i,t}) \) and \( Q_{i}^{'}(o_{i,t}, a_{i,t}, \overline{m}_{i,t}) \) respectively. \( \mathcal{SW}_{-i} \) can be estimated as follows:
\begin{align}
    &\mathcal{SW}_{-i} = \text{min}\left(\mathcal{O}_{max}, \text{max}\left(\mathcal{SW}^{'}_{-i}, \mathcal{O}^{-i}_{max} \right) \right),\\
    &\mathcal{SW}^{'}_{-i} = Q_{i}^{'}(\bm{0}, \bm{0}, \overline{m}_{i,t}^{0}) -  \gamma Q_{i}^{'}(\bm{0}, a^{0}_{i,t+1}, \overline{m}^{0}_{i, t+1}), \\
    &\overline{m}^{0}_{i, t} = \frac{1}{\mathcal{N}^i} \sum_{j \in \mathcal{N}(i)} m_{j,t}|_{ a_{j,t} = \mu^{'}(o_{j,t}, \overline{o}_{j,t}|_{o_{i,t} = 0}) }, \\
    &\overline{m}^{0}_{i, t+1} = \frac{1}{\mathcal{N}^i} \sum_{j \in \mathcal{N}(i)} m_{j,t+1}|_{a_{j,t+1} = \mu^{'}(o_{j,t+1}, \overline{o}_{j,t+1}|_{o_{i,t+1} = 0})},
\end{align}
where $\mathcal{O}_{max}$ and $\mathcal{O}^{-i}_{max}$ is the maximum value of $\sum_{i=0}^{N}{u}_{j,t}$ and $\sum_{j=0, j\neq i}^{N} u_{i,t}$ in $\mathcal{BI}$,  
\( \gamma \) is the discount factor, 
observation state $\bm{0}$ implies that no resources are allocated to VSP \( i \), and action $\bm{0}$ implies that there will be no adjustment for VSP's resource slices in this round. \( \overline{m}^{0}_{i, t} = \frac{1}{\mathcal{N}^i} \sum_{j \in \mathcal{N}(i)} m_{j,t}|_{ a_{j,t} = \mu^{'}(o_{j,t}, \overline{o}_{j,t}|_{o_{i,t} = 0}) }\),  \( a^{0}_{i,t+1} = \mu_{i}^{'}(\bm{0}, \overline{o}_{i,t+1})\) and \( \overline{m}^{0}_{i, t+1} = \frac{1}{\mathcal{N}^i} \sum_{j \in \mathcal{N}(i)} m_{j,t+1}|_{a_{i,t+1} = a^{0}_{i, t+1}}\)  can be calculated using \( \bm{o}_{t} \) and \( \bm{a}_{t} \). Besides, to compute the payment price for VSP \( i \), it is essential to estimate the utility of VSP \( i \) under such allocation; an MLP network is used for this estimation, denoted as \( f_{i}(\cdot) \), where $ U^{\ast}_{i,t} = f(o_{i,t}, a_{i,t})$, and VC pool updates as
\begin{align}
    &\mathcal{O}(t+1) = \mathcal{SW}^{\ast} = \sum_{i = 1}^{N}{f_{i}(o_{i,t},a_{i,t})}, 
\end{align}

In the second phase, the bank calculates the amount of tax the VSP must pay according to the currently known information. As stated in the equation (\ref{allocate}), by comparing the size of \(\mathcal{SW}^{\ast}\) and the sum of inverse bids \(x\), the allocation of resource slices in this round is determined. Noticeably, to meet constraint (\ref{C1}), the resource slice adjusted by the actor-network needs linear normalization, ensuring the total of each resource does not exceed the physical resource limit. 
Subsequently, according to equations (\ref{impact}) and (\ref{tax}), tax is calculated and the resource allocation and tax information are publicized.

In the third phase, the bank incorporates the financial report information \(\mathcal{R}\) collected in the current round, compounded with the intermediate information generated during the auction process of the round into the bill, i.e., into the replay buffer. The involved parameters in the auction are also updated through the replay buffer. With \(\alpha\) being estimated via the least squares method, its loss function is as follows
\begin{align}
    \nabla_{\alpha_{i}}\mathcal{L}(\alpha_{i}) = \mathbb{E}_{\mathcal{R}_{i} \sim \mathcal{BI}}\left[ (\frac{G_{i}}{U_{i}} - {\alpha}_{i})^2\right].
\end{align}
The loss function for the VoI network is formulated as
\begin{align}
    \nabla_{\phi_i}\mathcal{L}(\phi_{i}) = \mathbb{E}_{o_{i}, a_{i}, U_{i} \sim \mathcal{BI}}\left[ (f_{i}(o_{i}, a_{i}) - {U}_{i})^2\right].
\end{align}

For each agent's online actor network, the anticipated gradient update $\nabla_{\theta_i}  J(\mu_i)$ can be expressed as
\begin{align}
    \mathbb{E}_{\bm{o} \sim \mathcal{BI}, a_i \sim \mu_i} [\nabla_{\theta_i} \mu_i(o_{i},\overline{o}_{i}) \nabla_{a_i} Q_{i}(o_{i}, a_i,\overline{m}_{i})|_{a_i=\mu_i(o_{i}, \overline{o}_{i})}]. \notag
\end{align}
Herein, \(\mu_i\) is the policy of agent \(i\), \(Q_i^{\mu}\) is agent \(i\)'s online critic network, and the bill \(\mathcal{BI}\) is also considered as the experience replay buffer.
For the update of the online critic network, we typically employ Temporal Difference error to calculate the discrepancy between the target Q value and the current Q value, then employ back-propagation to update the parameters of the critic network. This can be depicted as:
\begin{align}
    \nabla_{\psi_i} \mathcal{L}(\psi_i) = \mathbb{E}_{\bm{o}, a, r, \bm{o}^{\dag} \sim \mathcal{BI}} [(Q_i(o_{i}, a_i,\overline{m}_{i}) - y)^2].
\end{align}
In this case, $\bm{o}^{\dag}$ is the next round observation state and \(y = r_i + \gamma Q_i'(o^{\dag}_{i}, a^{\dag}_i,\overline{m}^{\dag}_{i})|_{a_{i}^{\dag}=\mu'(o_{i}^{\dag}, \overline{o}^{\dag}_{i})}\) is the target Q-value.
Finally, the parameters of the target network are updated using a soft update mechanism. Specifically, for both the actor and critic network parameters, the update process can be expressed as:
\begin{align}
    \theta_i' = \tau \theta_i + (1 - \tau) \theta_i' ,\\
    \psi_i' = \tau \psi_i + (1 - \tau) \psi_i' ,
\end{align}
where \(\theta_i\) and \(\psi_i\) are the parameters of the online actor and critic networks, \(\theta_i'\) and \(\psi_i'\) are the parameters of the target actor and critic networks, and \(\tau\) is the ratio of the soft update. 
\begin{table}[htbp]
    \centering
    \caption{Parameter Setting}
    \renewcommand{\arraystretch}{1.2} 
    \setlength{\tabcolsep}{5pt} 
    \begin{tabular*}{0.40\textwidth}{@{\hspace{30pt}}l@{\hspace{50pt}}l@{\hspace{30pt}}}
    \hline
    Parameter  & Detail \\
    \hline
    Discount Factor $\gamma$ & 0.99  \\
    Learning Rate  & 0.001  \\
    Number of Episodes & 1000 \\
    Batch Size & 128 \\
    Replay Buffer Size & 10000 \\
    Soft Update Ratio $\tau$ & 0.005 \\
    \hline
    \end{tabular*}
    \label{tab: parameter}
\end{table}

\section{Simulation Results }
In the simulation section, we consider a multi-VSP network environment. The utility function of each VSP is set as
\begin{align}
 \label{utility-real}
  u(s_{0},s_{1},s_{2}) = \beta\left(1 - e^{-(\xi_{0}s_{0} + \xi_{1}s_{1} + \xi_{2}s_{2})}\right),
\end{align}
where $\beta \sim {U}(100, 200)$ and $\xi_{0},\xi_{1},\xi_{2} \sim {U}(0,10)$. $s_{0},s_{1},s_{2}$ denote the quantities of three distinct types of resource slices.
The profit of VSPs is formulated as indicated in equation (\ref{gross p}).
Without loss of generality, \(w \sim \mathcal{N}(\mu, \sigma^2)\) follows a normal distribution, unknown to the NSP. The total amount of all types of resources is normalized. In the simulation environment, the total number of VSPs is set between 10 and 50, and all algorithms are deployed under the same scenario with the same conditions. The involved additional parameters are consolidated in Table.\ref{tab: parameter}. The algorithms involved in the simulation are as follows:
\begin{itemize}
    \item \textbf{DVCG-MFMARL}(Proposed): A dual-currency VCG mechanism based on the MFMARL framework employs the dual-currency auction mechanism introduced in Section IV and the MFMARL to solve the slicing problem.

    \item \textbf{DVCG-MADDPG}: Similarly, it applies the auction mechanism presented in Section IV and utilizes the MADDPG framework to solve the slicing problem. Unlike the proposed algorithm, it uses joint actions and observed states as inputs for each agent network.

    \item \textbf{DVCG-MARL}: It also uses the auction mechanism introduced in Section IV, and employs the MARL framework to settle the slicing problem. It differs from the proposed algorithm by using only the observed states and actions of the corresponding VSP as inputs for each Agent's network.

    \item \textbf{QVCG}: The VCG mechanism proposed in \cite{8486341} under limited communication conditions, posits resource fragmentation into small quantized units. Each auction participant reports a bid vector instead of a utility function to achieve maximum social welfare allocation. This bid vector \(X = \{x_{1}, x_{2}, \ldots, x_{M}\}\) comprises of individual bids where \(x_{im} = u\left(\frac{m}{M}\right) - u\left(\frac{m-1}{M}\right)\), with \(\frac{1}{M}\) being the quantized resource unit. Relevant to our proposed scenario, due to three types of resources, a matrix of dimensions \(M \times M \times M\) is used instead of a bid vector.
\end{itemize}

\subsection{Convergence Evaluation}

\begin{figure}[htpb]
    \centering
    \includegraphics[width=3.4in]{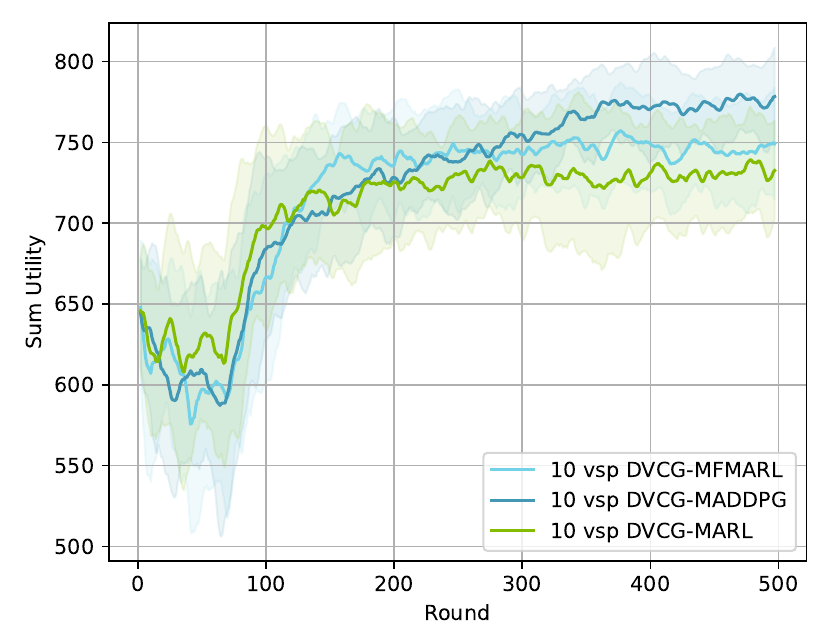}
    \caption{Comparative convergence of the total social welfare under different multi-agent reinforcement learning algorithms when the number of VSPs is 10.}
    \label{fig: Convergence1}
\end{figure}

\begin{figure}[htpb]
    \centering
    \includegraphics[width=3.4in]{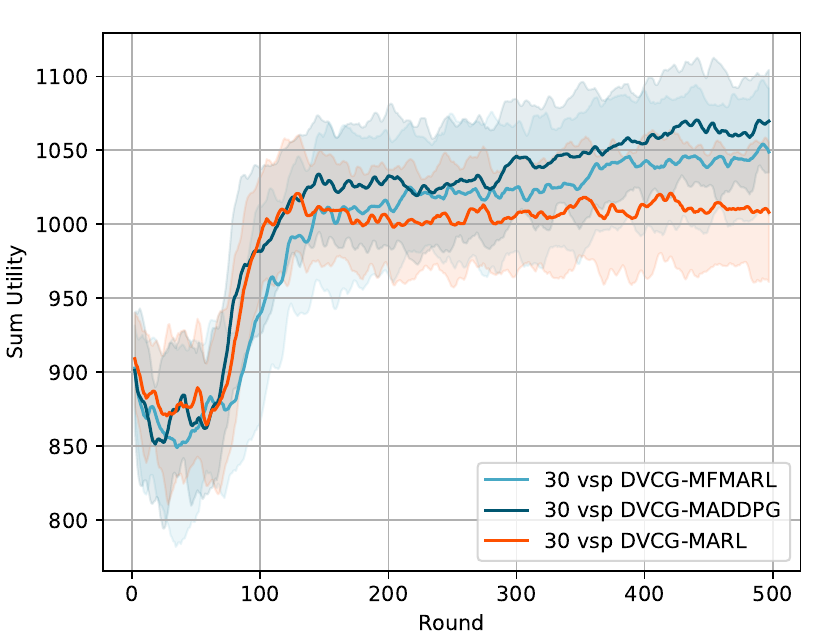}
    \caption{Comparative convergence of the total social welfare under different multi-agent reinforcement learning algorithms when the number of VSPs is 30.}
    \label{fig: Convergence2}
\end{figure}

We first show the convergence of the auction algorithm. As depicted in Fig. \ref{fig: Convergence1} and Fig. \ref{fig: Convergence2}, we present the convergence curve graphs for the DVCG-MFMARL, DVCG-MADDPG, and DVCG-MARL algorithms under 10 and 30 VSPs. QVCG is not an online learning algorithm, so its convergence is not discussed. Please note that in the above comparative simulations, VSP employs a bidding strategy to maximize net profits. Its strategy utilizes a DDPG architecture AC network as a substitute, with the reward function being the net profit of the current round. As illustrated, DVCG-MFMARL has a faster convergence rate compared to DVCG-MADDPG and DVCG-MARL, a consequence of DVCG-MFMARL having a smaller network size whilst witnessing more comprehensive global information. This advantage in the convergence speed of DVCG-MFMARL becomes more apparent as the number of intelligent agents increases. Simultaneously, DVCG-MADDPG has a higher total social welfare than DVCG-MFMARL, stemming from the inevitable information loss when DVCG-MFMARL simplifies observation and action dimensions using Mean Field Theory.

\subsection{Performance Evaluation}

Subsequently, we exhibit the cumulative social welfare achieved by four algorithms, wherein the auction was conducted for a total of 500 rounds. It is worth noting that for the QVCG algorithm, the precision of quantization, i.e., the size of the smallest resource unit, affects the attainable social welfare. Specifically, when the quantization precision is infinitely high, the allocation result theoretically approaches the optimal distribution. However, the dimension of the bid matrix also tends to be infinitely large. Moreover, the QVCG algorithm didn't consider the situation where bidders are not aware of their utility functions during its design. For the mechanism to be implemented, we conceded the QVCG algorithm, under the assumption that bidders are aware of parameters \(\alpha\), which is different from the settings of the other three online algorithms. For the QVCG algorithm, we introduced versions with quantization parameters of 0.01 and 0.05, respectively corresponding to different quantization accuracy.
\begin{figure}[htpb]
    \centering
    \includegraphics[width=3.4in]{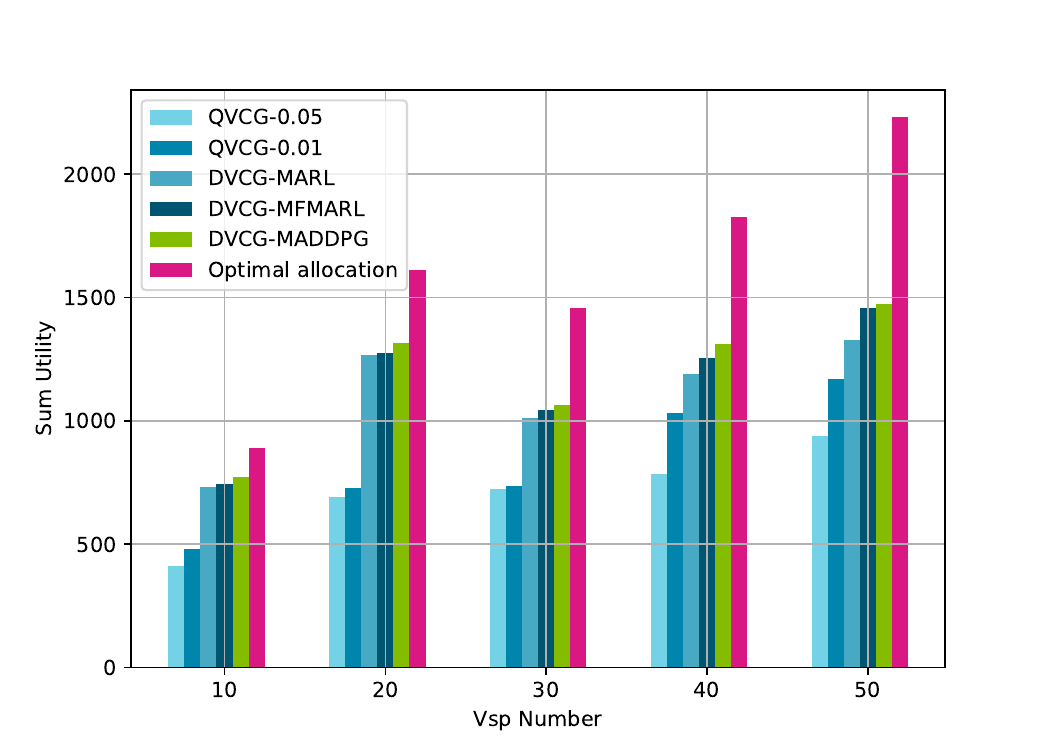}
    \caption{The relationship between the number of VSPs and the total social welfare under various resource slicing algorithms.}
    \label{fig: welfare}
\end{figure}

As shown in Fig. \ref{fig: welfare}, where the x-axis represents the number of intelligent agents and the y-axis represents the total social welfare, the DVCG-MADDPG, DVCG-MFMARL, and QVCG-0.01 algorithms more closely approach the optimal social welfare, outperforming the QVCG-0.01 and QVCG-0.05 algorithms. This performance gap narrows as the number of VSPs increases, mainly because a larger number of agents makes it more difficult for them to reach a tacit understanding. It should be noted that DVCG and QVCG mechanisms differ significantly in the amount of information exchange required, with DVCG-MFMARL demonstrating higher efficiency in this regard.

\begin{table}[h!]
\centering
\caption{Information exchange volume ( Float $\times 10^{6}$)}
\begin{tabular}{cccccc}
\hline
N & 10 & 20 & 30 & 40 & 50 \\
\hline
DVCG-MFMARL & 0.005 & 0.01 & 0.015 & 0.02 & 0.025\\
QVCG-0.05 & 0.08 & 0.16 & 0.24 & 0.32 & 0.40 \\
QVCG-0.01 & 10.0  & 20.0 & 30.0 & 40.0 & 50.0  \\
\hline
\end{tabular}
\label{Information Exchange Volume}
\end{table}

\begin{table}[h!]
\centering
\caption{Running time ($ \times 10^{3}$ S)}
\begin{tabular}{cccccc}
\hline
N & 10 & 20 & 30 & 40 & 50 \\
\hline
DVCG-MFMARL & 1.57 & 5.87 & 12.9 & 22.8 & 36.6\\
QVCG-0.05   & 0.19 & 1.82 & 3.68 & 9.35 & 30.4 \\
QVCG-0.01   & 3.40 & 33.6 & 117 & 225 & 1140 \\
\hline
\end{tabular}
\label{Time Complexity}
\end{table}

Table \ref{Information Exchange Volume} and Table \ref{Time Complexity} present the information exchange volume and runtime of three algorithms over 500 auction rounds with the number of VSPs ranging from 10 to 50. The information exchange volume measures the total amount of information exchanged between VSPs and NSPs, quantified in the number of floating-point values. Specifically, the DVCG mechanism requires the exchange of bid information in each auction round, so its information exchange volume is calculated as the number of VSPs multiplied by the number of rounds. In contrast, the QVCG mechanism requires each VSP to exchange a matrix of size \(M^{3}\) (where \(M\) represents the quantization precision), and this exchange occurs only once regardless of the number of auction rounds. The results show that in realistic scenarios (i.e., with a limited number of auction rounds), the information exchange volume of the DVCG mechanism is significantly lower than that of the QVCG-0.01 and QVCG-0.05 mechanisms. Additionally, the runtime of the DVCG mechanism falls between that of the two QVCG mechanisms with different precision levels.

\begin{figure}[htpb]
    \centering
    \includegraphics[width=3.4in]{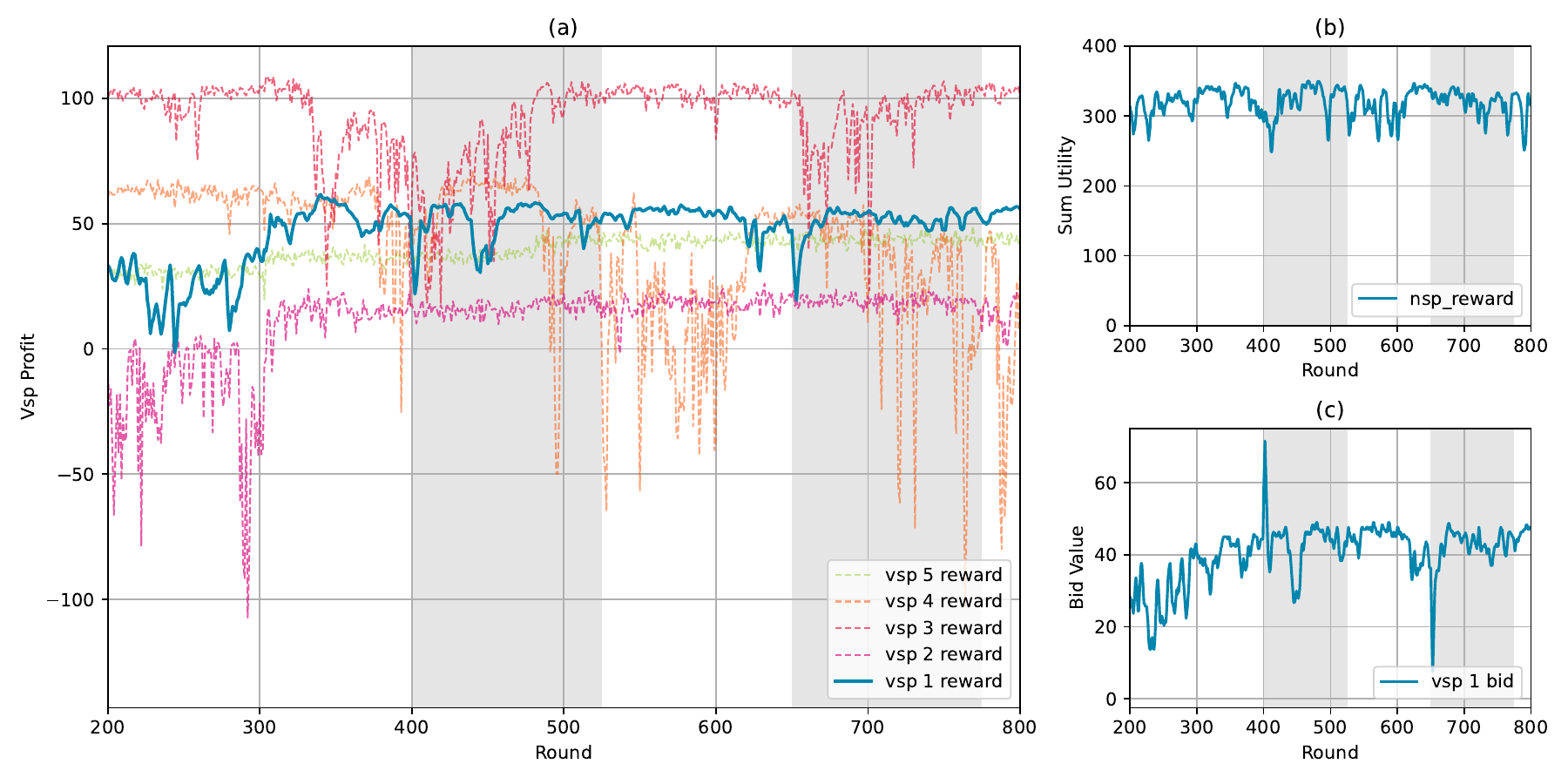}
    \caption{Verification of the DSIC properties in the DVCG-MFMARL Algorithm: a) Oscillations in VSP Profits; b) Variations in the total social welfare; c) Shifts in the bidding behavior of disruptive VSP.}
    \label{fig: DSIC}
\end{figure}

Furthermore, we elucidated the DSIC properties of the DVCG mechanism and its robustness in counteracting collusions among VSPs through our simulation. In Fig.\ref{fig: DSIC}, we demonstrate the DSIC characteristic of DVCG, where the VSP number is set to 5. The x-axis represents the number of auction rounds, while the y-axis records the net profit of the VSPs. Specifically, VSP 1 is a unique auction participant used to illustrate the DISC properties of the DVCG mechanism. It adjusts its bidding strategy in the 400th and 650th rounds. With the 400th round, VSP 1 commences inflating its bid above its true VoI. Starting with the 650th round, VSP 1 commences deflating its bid below the authentic VoI. It is observable that, regardless of the strategy employed, the net earnings for VSP 1 can never surpass those achieved when the true VoI is reported. However, it is worth noting that VSP 1's reporting of the bid can impact the net earnings of other VSPs. This necessitates the validation of the DVCG's robustness in resisting collusion.
\begin{figure}[htpb]
    \centering
    \includegraphics[width=3.4in]{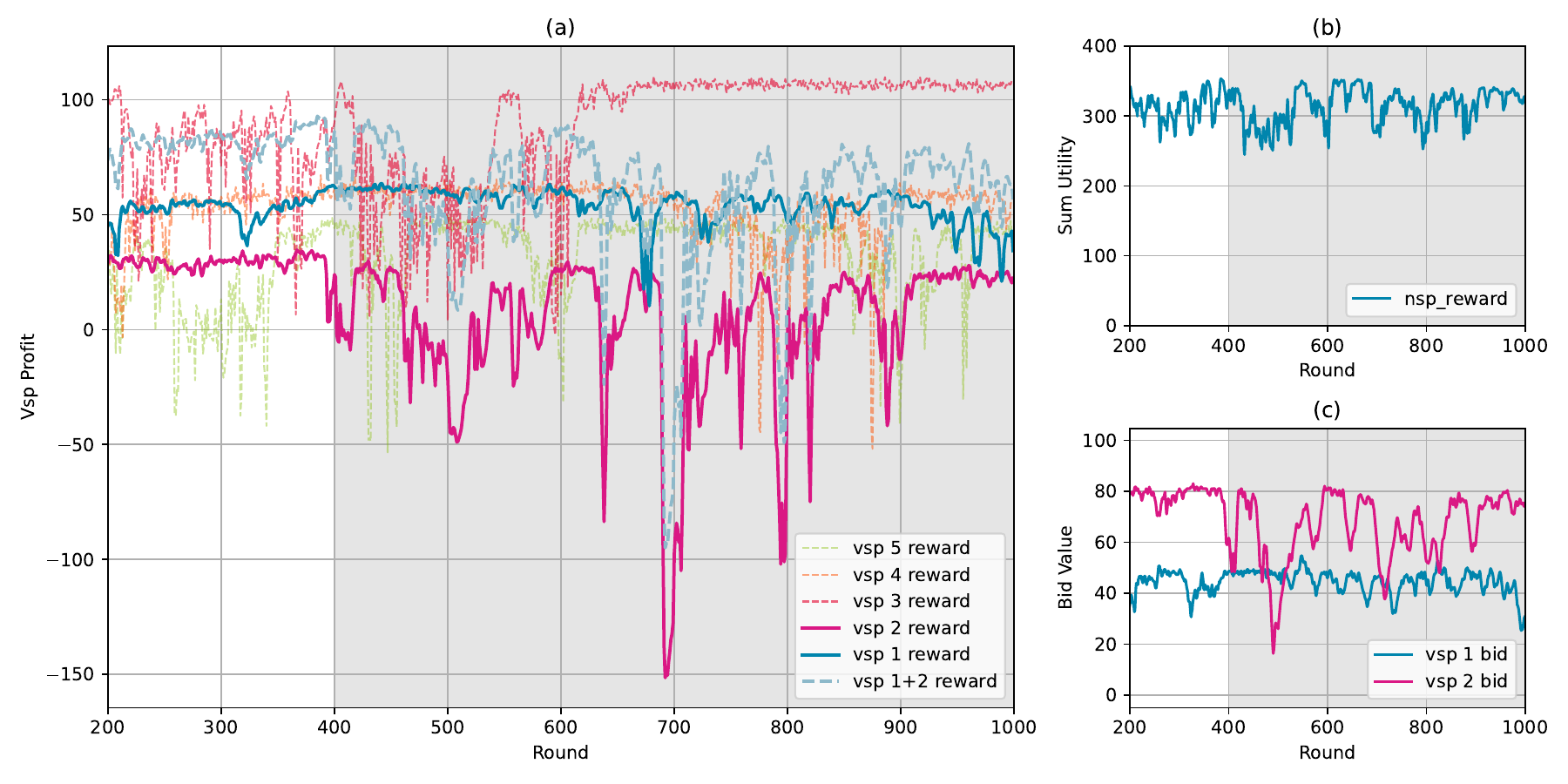}
    \caption{Evaluation of the DVCG-MFMARL algorithm's resilience against collusive behavior: a) Oscillations in VSP Profits; b) Variations in the total social welfare; c) Shifts in the bidding behavior of disruptive VSPs.}
    \label{fig: robustness1}
\end{figure}

With the same parameters, we demonstrated the robustness of the DVCG mechanism for collusion among auction participants. In this simulation, we selected two particular individuals from the five VSPs, denoted VSP 0 and VSP 1. VSP 0 and VSP 1 form a colluding set, which means that they use the same joint action agent, with the reward function of the bidding agents of the VSP 0 and VSP 1 set as 
\(r = P_{VSP0} + P_{VSP1}\).
We trained the VSP agents under the DVCG-FMMARL mechanism and illustrated the bids after the strategy of the VSP agents had converged, as depicted in Fig. \ref{fig: robustness1}. The blue and pink lines represent the net profits of VSP 0 and VSP 1. It can be observed that, even in collusion, the VSP 0 and VSP 1 agents still use the real bid as their dominant strategy after adequate exploration, proving the robustness of the DVCG mechanism against collusion.

\begin{figure}[htpb]
    \centering
    \includegraphics[width=3.4in]{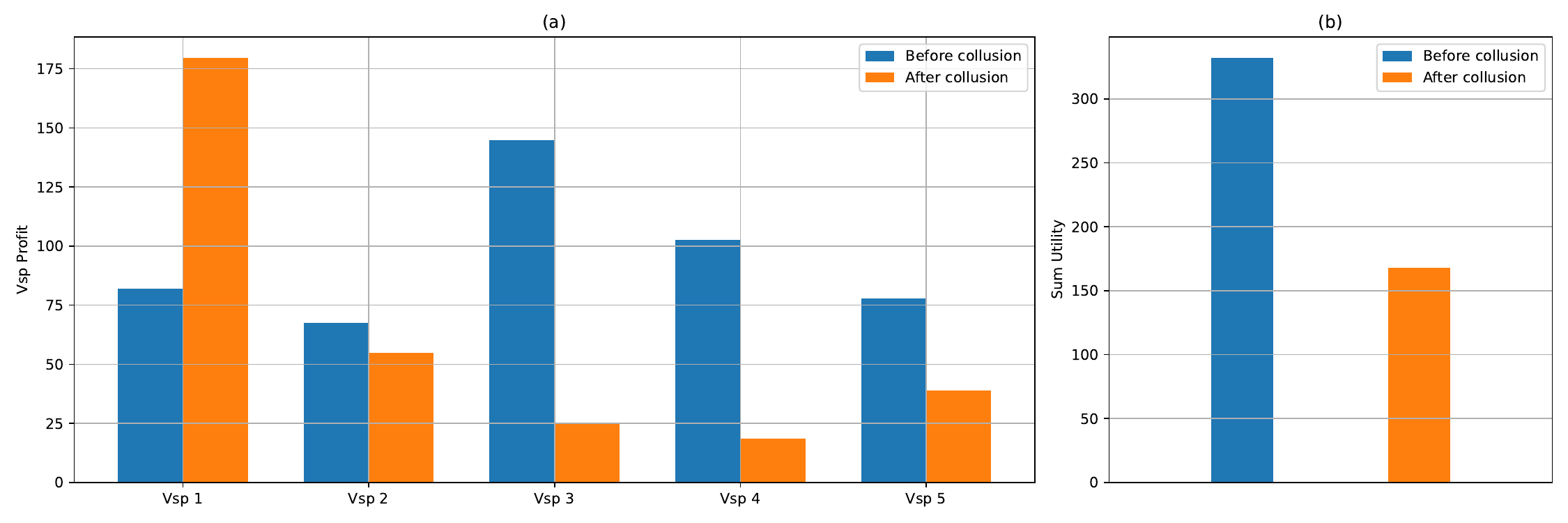}
    \caption{Assessment of the QVCG algorithm's vulnerability to collusive behavior: a) comparative analysis of VSP profits pre- and post-collusion; b) contrast of total social welfare before and after the collision.}
    \label{fig: robustness2}
\end{figure}
Simultaneously, we show the vulnerability of the QVCG mechanism when faced with collusion among VSPs. Due to the large dimensionality of the bid matrix under the QVCG mechanism, it is challenging to affirm its robustness against collusion through agent training. Therefore, we used an exhaustive search to demonstrate that under the QVCG mechanism, there is a colluding bidding scheme that, under the condition of collusion, increases the net profit of the colluding set (i.e., VSP 0 and VSP 1), as shown in Fig. \ref{fig: robustness2}.

\subsection{Ablation Study}

\begin{figure}[htpb]
    \centering
    \includegraphics[width=3.4in]{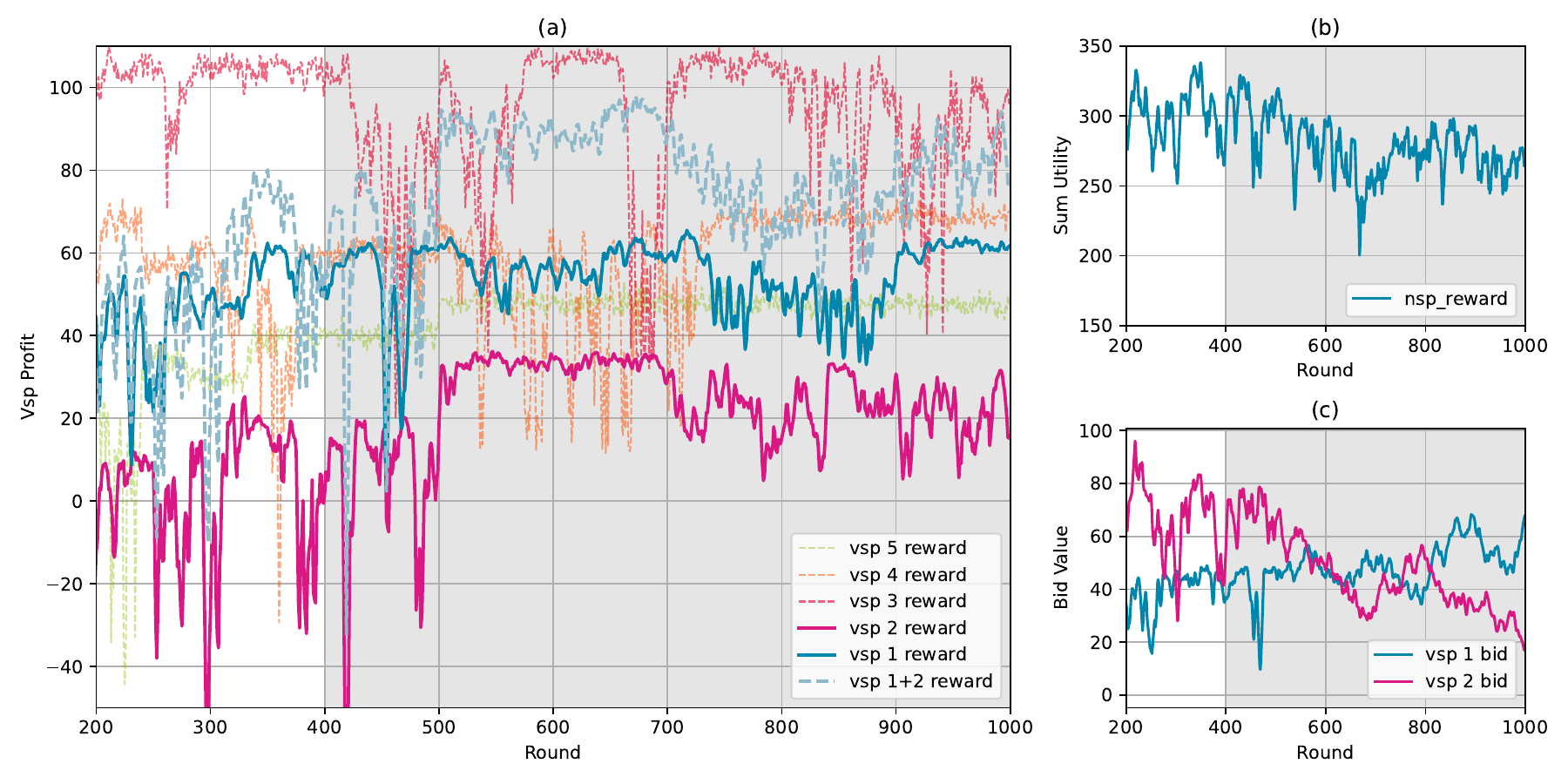}
    \caption{Assessment of the NVI-DVCG algorithm's vulnerability to collusive behavior: a) Oscillations in VSP Profits; b) Variations in the total social welfare; c) Shifts in the bidding behavior of disruptive VSPs.}
    \label{fig: ablation 1}
\end{figure}

Finally, we conducted ablation studies on the proposed mechanisms and introduced a variant mechanism based on DVCG-FMMRAL, termed NVI-DVCG, which eliminates the VC interest mechanisms from DVCG. The simulation setup is consistent with the simulation used to verify serial robustness. As shown in Fig. \ref{fig: ablation 1}, the NVI-DVCG mechanism significantly reduces the auction mechanism's resistance to serial stability compared to the DVCG mechanism. This diminished robustness allows VSP0 and VSP1 to collude, resulting in higher aggregate profits for the colluding parties but a decline in total social welfare.

\section{Conclusion}

This paper addresses the resource-slicing allocation problem in IoV by proposing a novel VoI-based dual-currency VCG auction mechanism. The mechanism achieves near-optimal social welfare improvement through optimized resource allocation without requiring knowledge of VSPs' revenue models. We theoretically and experimentally demonstrate that the mechanism satisfies DSIC properties, and can effectively mitigate potential collusion risks among bidders. 
Additionally, the mechanism significantly reduces the volume of information exchange per auction iteration while maintaining acceptable computational complexity, making it suitable for practical applications. 
In the future, we plan to extend our research to broader practical network scenarios and real-world optimization challenges. A representative focus will be solving multidimensional resource allocation problems in MIMO systems under complex constraints across frequency, temporal, spatial, and power domains, thereby broadening the horizons of this current discourse.

\bibliographystyle{IEEEtran}
\bibliography{main}


\end{document}